# All-Optical Spiking Neuron Based On Passive Micro-Resonator

JINLONG XIANG,[1] XUHAN GUO,[1,*] YIKAI SU,[1]

[1]*State Key Laboratory of Advanced Optical Communication Systems and Networks, Department of Electronic Engineering, Shanghai Jiao Tong University, Shanghai 200240, China*
*Corresponding author: guoxuhan@sjtu.edu.cn*

Neuromorphic photonics that aims to process and store information simultaneously like human brains has emerged as a promising alternative for the next generation intelligent computing systems. The implementation of hardware emulating the basic functionality of neurons and synapses is the fundamental work in this field. However, previously proposed optical neurons implemented with SOA-MZIs, modulators, lasers or phase change materials are all dependent on active devices and quite difficult for integration. Meanwhile, although the nonlinearity in nanocavities has long been of interest, the previous theories are intended for specific situations, e.g., self-pulsation in microrings, and there is still a lack of systematic studies in the excitability behavior of the nanocavities including the silicon photonic crystal cavities. Here, we report for the first time a universal coupled mode theory model for all side-coupled passive microresonators. Attributed to the nonlinear excitability, the passive microresonator can function as a new type of all-optical spiking neuron. We demonstrate the microresonator-based neuron can exhibit the three most important characteristics of spiking neurons: excitability threshold, refractory period and cascadability behavior, paving the way to realize all-optical spiking neural networks.

## 1. INTRODUCTION

Artificial intelligence (AI) has attracted a lot of attention lately because of its great potential to remarkably change almost every aspect of our daily life [1]. Applications of AI such as face recognition [2], intelligent virtual assistant [3] and neural machine translation [4] have already been widely used and brought great convenience to us. However, it takes a considerable amount of computational time to train the neural network for a specific task and AI operating on traditional computers consumes great power as well. In the von Neumann computing architecture, the computing unit and the memory are physically separated from each other, thus the machine instructions and data are serially transmitted between them. Consequently, the computational capacity is limited by the bandwidth and high energy consumption. Compared to the massively parallel processing of human brains, it's undoubtedly inefficient to train and simulate a neural network with software on a von Neumann computer.

Neuromorphic computing takes inspiration from human brains and aims to simultaneously process and store information as fast as possible, which provides a promising alternative for the next generation intelligent computing systems. It develops hardware mimicking the functions of neurons and synapses, and then combines them into suitably scaled neural networks. In the past few years, neuromorphic electronics, e.g., TrueNorth at IBM [5], TPU at Google [6] and SpiNNaker at the University of Manchester [7], have demonstrated impressive improvements in both energy efficiency and speed enhancement, allowing simultaneous operation of thousands of interconnected artificial neurons. However, electronic architectures face fundamental limits as Moore's law is slowing down [8]. Furthermore, the electronic communication links have intrinsic limitations on speed, bandwidth and crosstalk. Distinct from those properties of electronics, silicon photonics naturally has the advantages of low latency and low power consumption. Data is transmitted in the silicon waveguide at the speed of light and the associated energy costs currently are just on the order of femtojoules per bit [9]. Combined with the dense wavelength-division multiplexing (WDM), mode division multiplexing (MDM) and polarization division multiplexing (PDM) technology, the channel capacity continues to increase. Consequently, neuromorphic photonic systems could operate 6–8 orders-of-magnitude faster than neuromorphic electronics [10].

As shown in Figure 1, synapses and neurons are the basic building blocks of human brains, which perform the linear weight-and-sum operation and the nonlinear activation operation. To obtain a complicated brain-inspired optical chip, the hardware implementation of the fundamental function units is essential. Recently, a number of different concepts for neuromorphic computing have been proposed, including the reservoir computing [11-17], artificial neural networks (ANN) [18-25] and spiking neural networks (SNN) [26-39]. Utilizing the linear optics offered by the native interference property of light can implement completely passive optical matrix multiplication units for neural network applications in both free-space [21, 25] and the waveguide-based coherent circuitry [18]. Modulators combined with photodetectors [23, 24], SOA-MZIs [19, 38] and laser-cooled atoms with electro-magnetically induced transparency [25] are used to realize the nonlinear activation function in ANN. Meanwhile, lasers, e.g., micro-pillar lasers [36], two-section lasers with saturable absorber regions [26, 27], vertical-cavity surface-emitting lasers (VCSELs) [34, 35, 37, 40] and quantum dot (QD) lasers [39] are typically used to implement the functionality of spiking neurons in SNN. Owing to its unique optical properties in different states, phase change material (PCM) has emerged as an attractive alternative to provide in hardware both the basic integrate-and-fire functionality of neurons and the plastic weighting operation of synapses [28-31]. Moreover, the broadcast-and-weight architecture [33] has been proposed to unite well-defined neurons and synapses on the mainstream silicon photonics platform. However, the optical neurons proposed so far mainly rely on active devices, making it difficult to implement large-scale integrated neural networks. Although the PCM-based spiking neuron itself is passive, extended lasers are needed to generate the output pulses and the output signals are amplified off-chip to form the feedback links [31].

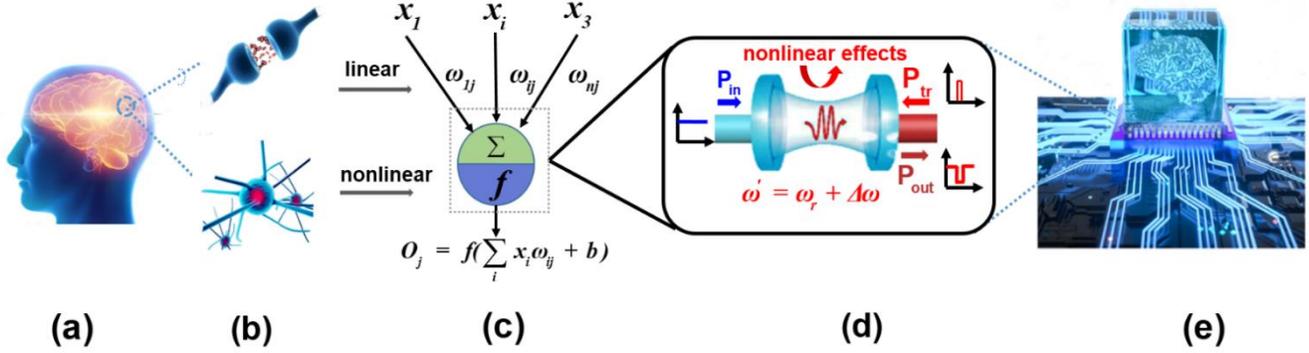

Figure.1. Schematics of optical neural networks. (a) The human brain; (b) The basic building blocks of brains: synapses and neurons; (c) The mathematic model of neurons, including the linear weight-and-sum operation and the nonlinear activation operation; (d) The microresonator-based optical spiking neuron, $\omega'$ is the effective resonance frequency, $\omega_r$ is the original resonance frequency and $\Delta\omega$ is the detuning of resonance frequency caused by the nonlinear effects; (e) The brain-inspired optical chip.

passive, extended lasers are needed to generate the output pulses and the output signals are amplified off-chip to form the feedback links [31].

In this paper, we report a new type of all-optical spiking neuron implemented with the nonlinearity of passive microresonators. When triggered with a strong enough perturbation signal, the microresonator will emit a negative pulse, which can be used to emulate the spiking behavior of neurons. To systematically study the interactions between different nonlinear effects, we propose for the first time a universe form of coupled mode theory (CMT) for all passive side-coupled microresonators. The self-pulsation and excitability behavior are successfully simulated to validate the presented model. Then, we demonstrate the microresonator-based neuron can exhibit the three most important characteristics of the spiking neuron: excitability threshold, refractory period and cascadability behavior. Moreover, we analyze the impact of Q factor on the cascadability of the proposed neuron. Finally, we provide the guideline to design a microresonator-based neuron for scalable neural networks.

## 2. COUPLED MODE THEORY OF MICRO-RESONATOR

The nonlinearity of microresonators has long been of interest as a simple mechanism to manipulate light with light. And the nonlinear phenomena have already been widely observed and carefully studied in different material platforms and various types of integrated resonators. All the self-pulsation, bistability and excitability behavior have been experimentally demonstrated in microrings [41-45], microdisks [46, 47] and two-dimensional photonic crystal (PhC) cavities in InP platform [48, 49]. And the silicon two-dimensional PhC cavities can exhibit bistability [50-52] and self-pulsation [53-56] behaviors. Besides, bistability has also been observed in racetrack resonator [57] and silicon one-dimensional PhC cavity [58, 59]. While excitability behavior hasn't been demonstrated in silicon PhC nanocavities.

The basic principle of nonlinear behaviors for Silicon-On-Insulator (SOI) microresonators is [44]: two-photon absorption (TPA) effect first generates free carriers, which are then able to absorb light by free carrier absorption (FCA) effect. Besides, the presence of free carriers leads to a blueshift in the resonance wavelength by free carrier dispersion (FCD) effect. Moreover, surface state absorption takes place everywhere at the silicon-silica interface, and meanwhile some light is lost due to the surface scattering and radiation loss. Therefore, the absorbed optical energy is lost mainly by heat. Owing to the thermo-optic effect, the heat induces a redshift in the resonance wavelength. Typically, the free carriers relax at least one order of magnitude faster than the heat. Consequently, this difference between the timescale of the fast free carrier dynamics and the slow heating effects results in the self-pulsation, bistability and excitability phenomena in microresonators.

Previously, different forms of CMT have been proposed to analyze the nonlinearity in silicon microrings [42-44], microdisks [46, 47] and two-dimensional silicon PhC cavities [53, 55, 56, 60]. However, these CMT equations are intended for specific situations, e.g., self-pulsation in microrings. The lack of systematic studies in the nonlinear excitability behavior of nanocavities including the silicon PhC cavities is pressing. In this paper, we first propose, to the best of our knowledge, a universal CMT model for all passive side-coupled microresonators. Considering the most common situation where a passive microresonator couples with only one waveguide, the nonlinear dynamics of the microresonator can be described as followed [44, 46, 51, 61, 62]:

$$\frac{da_+}{dt} = \left[ j(\omega' - \omega_+) - \frac{1}{2\tau_{+,total}} \right] a_+ + \kappa_+ S_+. \quad (1)$$

$$\frac{da_-}{dt} = \left[ j(\omega' - \omega_-) - \frac{1}{2\tau_{-,total}} \right] a_- + \kappa_- S_-. \quad (2)$$

$$\frac{dN}{dt} = -\frac{N}{\tau_{fc}} + \frac{\Gamma_{FCA}\beta_{Si}c^2}{2\hbar\omega V_{FCA}n_g^2} \left( |a_+|^4 + 4|a_+|^2|a_-|^2 + |a_-|^4 \right). \quad (3)$$

$$\frac{d\Delta T}{dt} = -\frac{\Delta T}{\tau_{th}} + \frac{\Gamma_{th}P_{abs}}{\rho_{Si}C_{p,Si}V_{cavity}}. \quad (4)$$

Where $a_\pm$ is the complex amplitude of the forward and backward propagation mode respectively and $|a_\pm|^2$ stands for the corresponding mode energy in the microresonator. $S_\pm$ represents the complex amplitude of the pump light and the perturbation signal injected in the opposite direction. $\omega_\pm = \frac{2\pi c}{\lambda_\pm}$ is the frequency of input light in the waveguide, $\omega' = \omega_r + \Delta\omega_i$ denotes the shifted resonance frequency of the microresonator, where $\omega_r = \frac{2\pi c}{\lambda_r}$ is the original resonance frequency and $\Delta\omega_i$ is the total frequency shift caused by all nonlinear effects. $\kappa_\pm = \sqrt{\frac{1}{\tau_{\pm,in}}}$ is the coupling coefficient between the waveguide and the microresonator. $N$ is the concentration of free carriers in the microresonator and $\Delta T$ is the mode-averaged temperature difference with the surroundings. $\tau_{th}$ is the relaxation time for temperature and $\tau_{fc}$ is the effective free-carrier decay rate accounting for both recombination and diffusion. $\beta_{Si}$ is the constant governing TPA, $\rho_{Si}$ is the density of silicon, $C_{p,Si}$ is the thermal capacity of silicon and $V_{cavity}$ is the volume of the microresonator. $n_g$ is the group index, generally, the dispersion is neglected and $n_g = n_{Si}$, $n_{Si}$ is the refractive index of bulk silicon. $V_\alpha$ and $\Gamma_\alpha$ denote the effective mode volume and confinement coefficient (see Supplement 1 for additional details).

The total loss for each cavity mode is:

$$\frac{1}{\tau_{i,total}} = \frac{1}{\tau_{i,lin}} + \frac{1}{\tau_{i,v}} + \frac{1}{\tau_{i,in}} + \frac{1}{\tau_{i,TPA}} + \frac{1}{\tau_{i,FCA}}. \tag{5}$$

where $\frac{1}{\tau_{i,lin}}$ is the linear absorption loss, $\frac{1}{\tau_{i,in/v}}$ represents the in-plane waveguide coupling loss and the vertical radiation loss respectively, given by $\frac{1}{\tau_{i,in/v}} = \frac{\omega}{Q_{in/v}}$; $\frac{1}{\tau_{i,TPA}}$ and $\frac{1}{\tau_{i,FCA}}$ are the loss due to TPA and FCA respectively. The mode-averaged TPA loss rates are:

$$\frac{1}{\tau_{+,TPA}} = \Gamma_{TPA} \frac{\beta_{Si} c^2}{n_g^2 V_{TPA}} \left( |a_+|^2 + 2|a_-|^2 \right). \tag{6}$$

$$\frac{1}{\tau_{-,TPA}} = \Gamma_{TPA} \frac{\beta_{Si} c^2}{n_g^2 V_{TPA}} \left( |a_-|^2 + 2|a_+|^2 \right). \tag{7}$$

The mode-average FCA loss rate is defined as:

$$\frac{1}{\tau_{i,FCA}} = \frac{c}{n_g} \left( \sigma_e + \sigma_h \right) N. \tag{8}$$

From the Drude model, the absorption cross-section for electrons and holes $\sigma_{e/h}$ is

$$\sigma_{e/h} = \frac{e^2}{c n_g \omega_r^2 \varepsilon_0 m^*_{e/h} \tau_{relax,e/h}}. \tag{9}$$

Here $e$ is the electron charge, $\tau_{relax,e/h}$ is the relaxation time of free carriers, $\varepsilon_0$ is the permittivity of free space and $m^*_{e/h}$ is the effective mass of carriers.

The total detuning of the resonance frequency of the microresonator $\Delta\omega_i$ can be expressed by:

$$\frac{\Delta\omega_i}{\omega_r} = -\frac{\Delta n_i}{n_{Si}} = -\left( \frac{\Delta n_{i,Kerr}}{n_{Si}} + \frac{\Delta n_{i,FCD}}{n_{Si}} + \frac{\Delta n_{i,th}}{n_{Si}} \right). \tag{10}$$

The detuning due to the Kerr effect is

$$\frac{\Delta n_{+,Kerr}}{n_{Si}} = \frac{c n_2}{n_g V_{Kerr}} \left( |a_+|^2 + 2|a_-|^2 \right). \tag{11}$$

$$\frac{\Delta n_{-,Kerr}}{n_{Si}} = \frac{c n_2}{n_g V_{Kerr}} \left( |a_-|^2 + 2|a_+|^2 \right). \tag{12}$$

where the effective mode volume for Kerr effects $V_{Kerr} = V_{TPA}$ and $n_2$ is the Kerr coefficient.

The detuning due to the FCD effect can be expressed by

$$\frac{\Delta n_{i,FCD}}{n_{Si}} = -\frac{1}{n_{Si}} \frac{dn_{Si}}{dN} N = -\frac{1}{n_{Si}} \left( \zeta_{i,e} + \zeta_{i,h} \right) N. \tag{13}$$

Here the material parameter with units of volume $\zeta_{i,e/h}$ is defined as

$$\zeta_{i,e/h} = \frac{e^2}{2 n_{Si} \omega_i^2 \varepsilon_0 m^*_{e/h}}. \tag{14}$$

Finally, the detuning due to the thermal dispersion is:

$$\frac{\Delta n_{i,th}}{n_{Si}} = -\frac{1}{n_{Si}} \frac{dn_{Si}}{dT} \Delta T. \tag{15}$$

The total absorbed power is given by:

$$P_{abs} = P_{abs,lin} + P_{abs,TPA} + P_{abs,FCA}. \tag{16}$$

where

$$P_{abs,lin} = \frac{1}{\tau_{lin}} \left( |a_+|^2 + |a_-|^2 \right). \tag{17}$$

$$P_{abs,FCA} = \frac{1}{\tau_{FCA}} \left( |a_+|^2 + |a_-|^2 \right). \tag{18}$$

$$P_{abs,TPA} = \Gamma_{TPA} \frac{\beta_{Si} c^2}{n_g^2 V_{TPA}} \left( |a_+|^4 + 4|a_+|^2 |a_-|^2 + |a_-|^4 \right). \tag{19}$$

## 3. NONLINEAR BEHAVIORS OF MICRORESONATORS

To validate the universal CMT-equations presented above, we simulate the nonlinear self-pulsation and excitability behaviors of microresonators taking the most common microrings and nanobeams as examples (parameters used in the simulation are provided in Supplement 1). Since it takes a relatively long time for the microring to reach its steady state from scratch, all the simulation results for microrings begin with the steady state in demonstration. In simulation the pump light and the perturbation signal have the same wavelength, and the backscattering is neglected. As a consequence, there exists no interference effect due to the coupling between the pump light and the trigger light. Moreover, the dynamics of the microresonator will be independent of the phase of the perturbation signal. The time width of the trigger pulse is 10ns and 1ns for the microring and the nanobeam respectively. Besides, the maximum power of the perturbation signal is set to be the corresponding pump power.

The optical energy in the microresonator generates both heat and free carriers, while the thermo-optic effect and the FCD effect have an opposite influence on the resonance frequency shift. Due to the difference in the timescale between the slow relaxation process of heat and the fast free carrier generation and absorption dynamics, microresonators will exhibit self-pulsation behavior, as shown in Figure 2. The mode energy in the microcavity, the temperature difference between the microcavity and the surroundings, the concentration of free carriers, and the power of output light are all evolving periodically with time.

Excitability was behaviorally defined by three main criteria: (i) an unperturbed system rests at a single stable equilibrium; (ii) an external perturbation above the excitability threshold triggers a large excursion from its equilibrium; (iii) the system then settles back to the attractor in what is called the refractory period, after which the system can be excited again [63]. In fact, if the input power is below but very close to the self-pulsation power, the microresonator can be excited to emit a negative output pulse when applied a strong enough perturbation signal. As shown in Figure 3, the microresonator initially rests at its steady state and the output power remains constant. When triggered with a strong enough perturbation signal, the microresonator will be kicked out from its equilibrium. All the mode

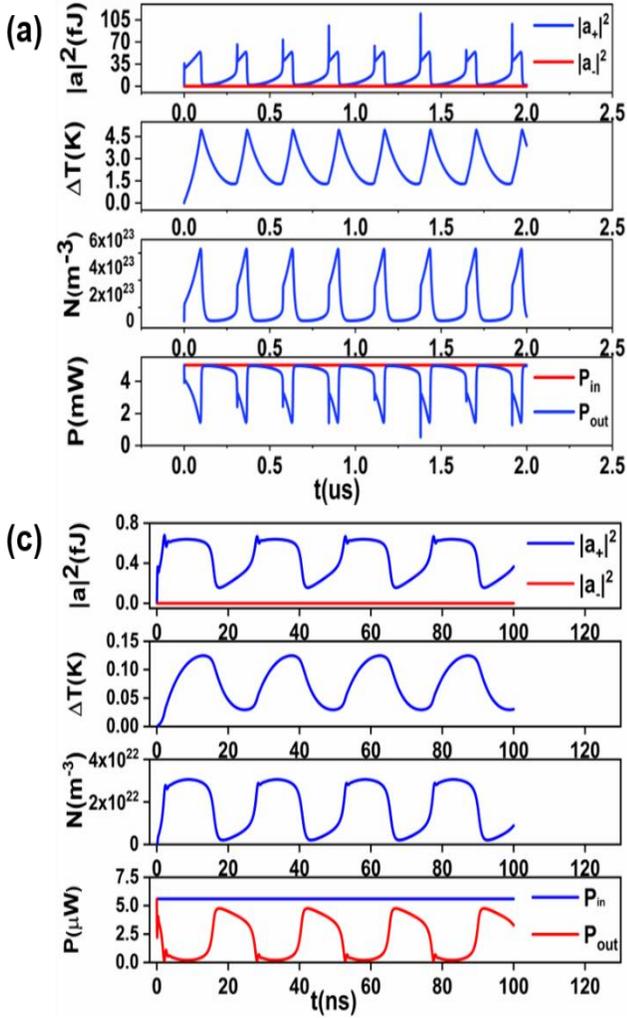
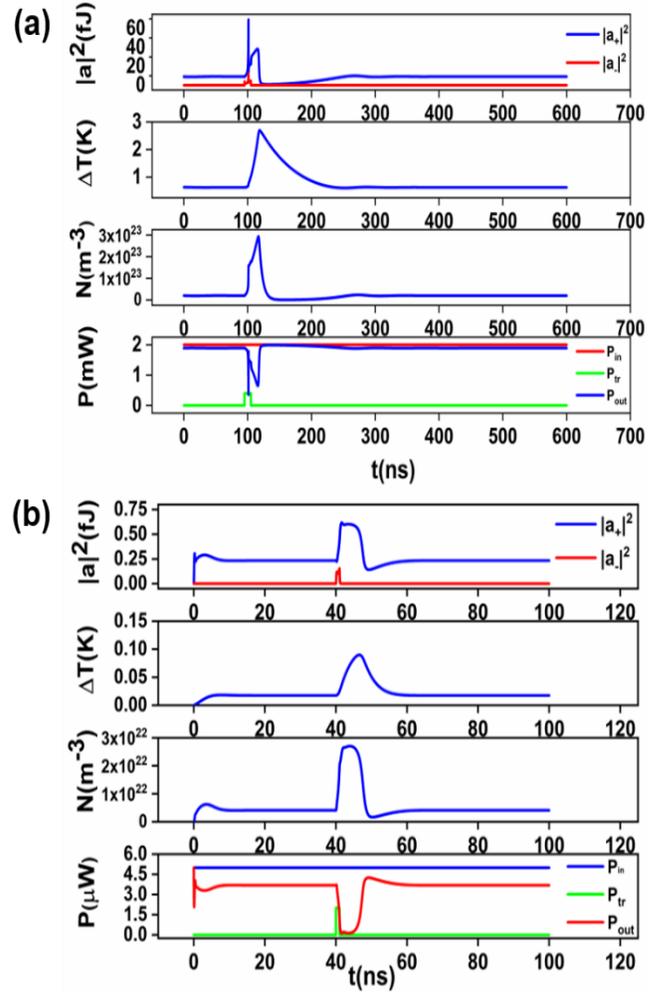

Figure.2. Schematics of the self-pulsation behavior of microresonators. (a) For microring: the detuning $\delta_\lambda = -20\,pm$ and $P_{in} = 5\,mW$; (b) For nanobeam: $\delta_\lambda = -12\,pm$ and $P_{in} = 5.6\,\mu W$.

Figure.3. Schematics of the excitability behavior of microresonators. Under proper pump conditions, the microresonator can be excited with a strong enough perturbation signal. (a) For microring: $\delta_\lambda = -20\,pm$, $P_{in} = 2\,mW$ and $P_{tr} = 0.4\,mW$; (b) For nanobeam: $\delta_\lambda = -13\,pm$, $P_{in} = 5\,\mu W$ and $P_{tr} = 2\,\mu W$

cavity energy, the temperature difference between the resonators and their surroundings and the concentration of free carriers go through a sudden increase and then slowly recover to their steady states, resulting in the appearance of a negative pulse in output power. Moreover, as long as the power of the trigger signal is above the threshold value, the shape of the negative output pulse will keep the same. The simulation results agree quite well with previous studies [44, 55], which guarantees the correctness of the proposed universal CMT-equations.

## 4. OPTICAL SPIKING NEURON BASED ON PASSIVE MICRORESONATOR

Although the microresonator emits a negative pulse when excited with a perturbation signal, it can still be used to mimic the basic functions of a spiking neuron. In SNN, the information is encoded and transmitted in the form of spikes. It doesn't matter what the shape of the spike is, and whether the spike is positive or negative also makes no difference to the operation of the whole SNN. Here, we explicitly point out that the passive microresonator can serve as an all-optical spiking neuron.

. The spiking neuron has three most important characteristics: (i) excitability threshold behavior: The spiking neuron will only emit an output pulse when the power of the perturbation signal exceeds a certain threshold. Otherwise, the neuron will not respond and remain silent. (ii) refractory period behavior: After excitation, the neuron needs to relax to its steady-state before it can be triggered once again, and the recovery time needed is called the refractory period. (iii) cascadability behavior: The output spike of the previous neuron should be strong enough to excite the next neuron, which is the foundation of forming a multi-layer spiking neural network. Next, we provide a detailed exposition of the major characteristics of the nanobeam-based spiking neuron. (The microring-based neuron can exhibit the same behaviors as nanobeams, and the simulation results are provided in Supplement 1)

A. Excitability threshold

In the ANN, the nonlinear unit is a necessary part for learning tough tasks. Without the nonlinear activation function, the output of ANN will only be the simple linear combination of the input information. Consequently, the ANN turns out to be the primitive perceptron machine. It's the nonlinear unit that makes it possible for ANNs to approximate any functions, thus realizing complicated tasks such as image classification, speech recognition and so on. The sigmoid function and tanh function are usually adopted in the ANN. While the information in SNN is encoded in the timing or rate information of spikes, which are assumed to be all-or-none binary events. The spikes are used for interneuron communication and carry information through the sheer fact of their appearance, whereas their amplitude and width are neglected. Consequently, the condition on which the spiking neuron can be triggered to emit an output spike, or more exactly, the excitability threshold behavior, directly determines the performance of the SNN.

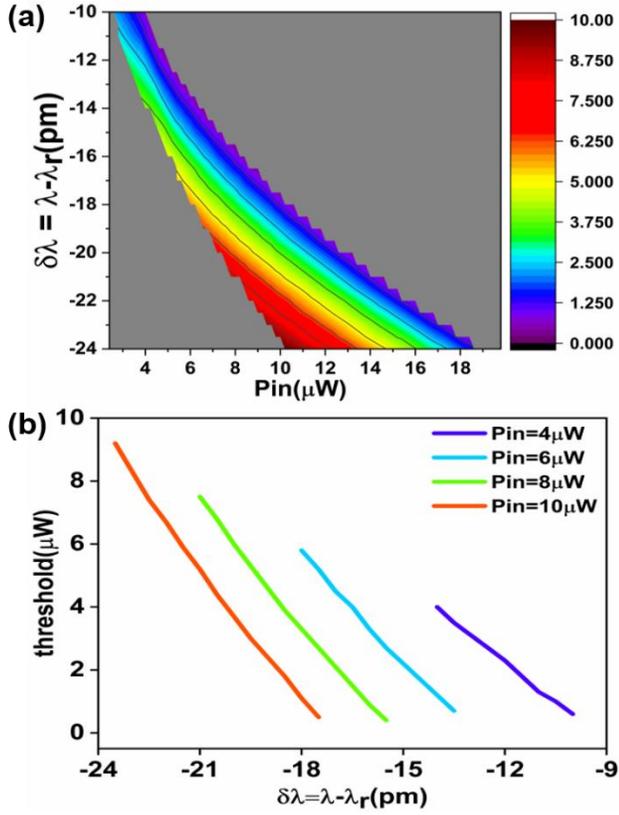
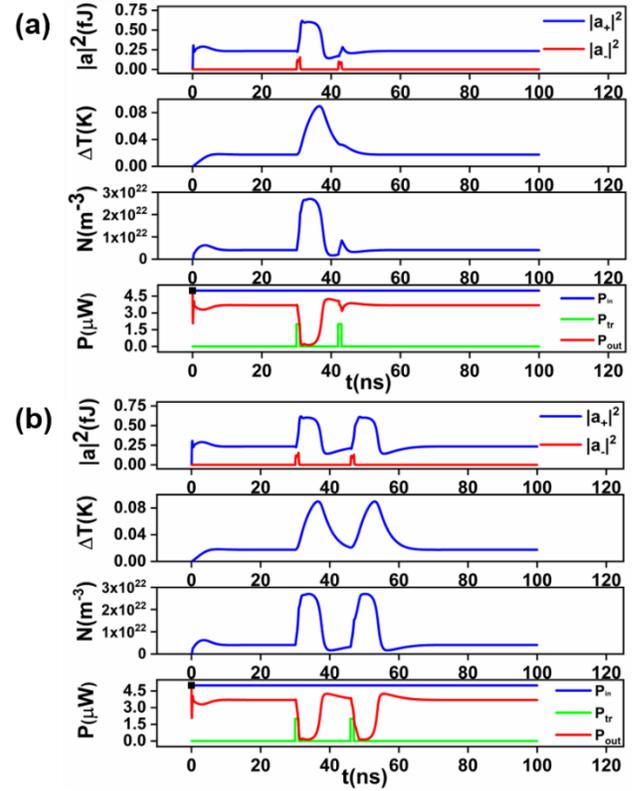

Figure.4. Schematics of the distribution of the excitability threshold of nanobeam over different pump power $P_{in}$ and wavelength shift $\delta_\lambda$.

Figure.5. Schematics of the refractory period behavior of nanobeam. For $\delta_\lambda = -13\,pm$, $P_{in} = 5\,\mu W$, $P_{tr} = 2\,\mu W$, (a) the nanobeam keeps silent to the second trigger pulse when the time interval is 12ns, (b) but it can be excited once again when the time interval increases to 16ns.

The microresonator-based spiking neuron exhibits obvious threshold behavior. Figure 4 and Figure S2 present the distribution of the excitability threshold over different input power and resonance shift. As shown in Figure 4(a) and Figure S2(a), the microring and the nanobeam have similar distribution characteristics. However, the excitable condition for nanobeams is more demanding than microrings, owing to its higher Q/V ratio. It's worth mentioning that the microresonator will exhibit self-pulsation or bistability behavior at the upper right gray region. Moreover, it can be excited at the lower left gray region, but the trigger power needs to be higher than the pump power, which is not discussed in this paper. According to Figure 4(b) and Figure S2(b), as the wavelength shift increases, the excitability threshold power of microresonators increase gradually. On the contrary, it becomes lower when pumped with higher input power. In addition, the excitability threshold is more sensitive to the wavelength shift than the pump power.

B. Refractory Period

After an excitation, the spiking neuron will be insensitive to the second trigger signal until recovering to its steady state. When scaled to a multi-layer SNN, the refractory period will directly determine the operation speed of the whole network.

The microresonator-based optical neuron has this property as well. As shown in Figure 5 and Figure S3, if the interval between the two trigger pulses is shorter than the refractory period, the microresonator will not respond to the second perturbation signal. For $\delta_\lambda = -13\,pm$, $P_{in} = 5\,\mu W$, $P_{tr} = 2\,\mu W$, when the time interval is 12ns, the nanobeam stays silent to the second trigger pulse. However, when the time interval increases to 16ns, the nanobeam emits an output spike once again, which is the direct evidence of the refractory period behavior.

The effective relaxation time of free carrier $\tau_{fc}$ and the relaxation time of temperature $\tau_{th}$ are two key factors deeply influencing the excitability behavior of microresonators. The former determines the shape of the negative pulse, while the latter determines how fast the microresonator can be excited. Beneficial from the shorter thermal relaxation time, the refractory period for nanobeams ($\sim 20\,ns$) is approximately one eighth of that for microrings ($\sim 160\,ns$).

C. Cascadability Behavior

To provide the required learning capacity for complicated artificial intelligence applications such as computer vision and medical diagnosis, a multi-layer neural network is the inevitable choice, which means the cascadability behavior of optical neurons is quite necessary. In order to simulate the cascadability behavior of the proposed optical neuron, two

## TABLE 1. Performance Comparison of Optical Neuron

| | Type | Power | Operation speed | Cascadability | Integration | Passive/active |
|---|---|---|---|---|---|---|
| Microring(proposed) | SNN | mW | sub-µs | Yes | High | passive |
| Nanobeam(proposed) | SNN | $\mu W$ | ns | Yes | High | passive |
| SOA-MZI [19, 38] | ANN | mW | sub-ns | Yes | Low | active |
| Laser cooled atoms [25] | ANN | sub-mW | ms | Yes | Low | active |
| Modulator [23, 24] | ANN | mW | ns | Yes | medium | active |
| Integrated laser [26, 27, 35-37, 40] | SNN | $\mu W/mW$ | sub-ns/ns | Yes | medium | active |
| PCM [29, 31] | SNN | nJ | ns | No | medium | active |

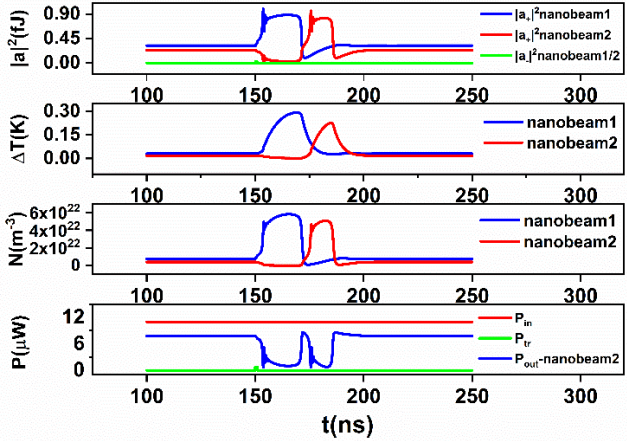

Figure.6. Schematics of the cascadability behavior of microresonators. For $\delta_\lambda = -18.5\,pm$, $P_{in} = 11\mu W$, $P_{tr} = 0.75\mu W$, the perturbation signal excited the first nanobeam and its negative output pulse triggers the second nanobeam.

microresonators with the same conditions are cascaded. The perturbation signal is only applied to the first microresonator, and the output signal of the first neuron directly serves as the input signal of the second neuron. The simulation results are presented in Figure 6 and Figure S3. For $\delta_\lambda = -18.5\,pm$, $P_{in} = 11\mu W$, $P_{tr} = 0.75\mu W$, the perturbation signal excites the first nanobeam and its negative output pulse triggers the second nanobeam. All the physical parameters of the second microresonator including the mode-averaged energy in the cavity, the temperature difference with the surroundings and the free carrier concentration evolve in the same way as the first microresonator after a certain latency. Consequently, there exist two negative pulses at the output of the second microresonator. The first negative pulse corresponds to the output of the first neuron, and the second pulse is the triggered output of the second neuron, which is the direct evidence for the cascadability behavior of the microresonator-based spiking neuron.

## 5. DISCUSSION

We analyze the performance of the microresonator-based neuron: the power consumption and the operation speed are determined by the pump power and the refractory period respectively. As discussed above, the microring-based neuron can be pumped and excited with serval milliwatts. Owing to the ultra-high Q/V ratio, the light-matter interaction in the nanobeam is much stronger than that in the microring. Consequently, the nanobeam-based neuron can work under serval microwatts, three orders-of-magnitude lower than the microring. Moreover, since the refractory period is mainly limited by the relaxation time of heat, the nanobeam-based neuron can operate more than eight times faster than the microring-based neuron. We emphasize that the microresonator-based neuron can operate at the speed of GHz, six orders-of-magnitude faster than the timescale of biological neural networks: kHz. Table 1 depicts the performance comparison between our proposed optical neuron with previous work. The microresonator-based neuron can process information passively, which is extremely difficult for other types of neuron to achieve. Moreover, they can be easily fabricated in photonic integrated circuits together with other optical components in standard CMOS process, thus making large-scale SNNs available. Combined with the PCM-implemented synapse in [30], an all-optical passive spiking neural network can be obtained.

Furthermore, we investigate the impact of the Q factor on the distribution of the cascadability behavior of the nanobeam-based neuron. Since ultra-high Q of $3.6\times10^5$ on SiO2 claddings and $7.2\times10^5$ on air claddings have been experimentally demonstrated [64], we keep the $Q_v/Q_{in}$ ratio constant and increase the Q value from original 114400 to 171600 and 228800 successively. The simulation results are depicted in Figure 7. As the value of Q increases from left to right, relatively larger wavelength detuning and higher pump power are needed for the nanobeam-based neuron to demonstrate the cascadability behavior. It can be attributed to the high temperature sensitivity of the resonance frequency resulting from the ultra-high Q/V ratio.

Finally, we give the general guideline to design a practical microresonator-based neuron: *(i) Choose the type of microresonator.* Compared with microrings, the nanobeam-based neuron can work with faster speed and much lower power. However, owing to the ultra-high Q/V ratio, nanobeams can be more sensitive to the temperature variation of the surroundings, affecting the stability of the whole system. Moreover, the condition for the cascadability of nanobeams is more critical; *(ii) Design the quality factor $Q_v$ and $Q_{in}$ of the microresonator.* Generally, a higher Q can enhance the light-matter interaction, thus making the microresonator exhibit the expected nonlinear behaviors. While as discussed above, a lower Q is preferred to more easily achieve the cascadability with lower threshold; *(iii) Choose proper working region: wavelength detuning and pump power.* Combining the PCM-based synapse and the proposed optical neuron, a two-layer SNN as demonstrated in [31] can be easily implemented, where just the excitability of neurons is needed. However, to realize complex multi-layer neural networks, the neuron must be working in the region where the microresonator can exhibit cascadability behavior.

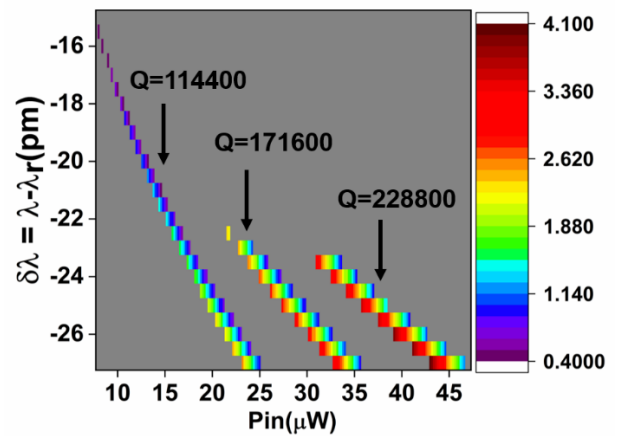

Figure.7. Schematic of the influence of Q factor on the distribution of the cascadability behavior of nanobeam-based neuron. From left to right, $Q = 114400, Q = 171600$ and $Q_v = 228800$ respectively.

## 6. CONCLUSION

In summary, we creatively utilize the nonlinearity of microresonators to mimic the functions of spiking neurons. The microresonator-based neuron can demonstrate the excitability threshold behavior, refractory period behavior and cascadability behavior, which is of great importance for the spiking neuron. We emphasis that the proposed neuron is completely passive and can be easily fabricated with standard CMOS process. Moreover, we report a universal CMT model that can be used to analyze the nonlinear dynamics in all side-coupled microresonators, especially for the excitability of silicon PhC cavities. Combining the PCM-based synapse and the microresonator-based neuron, multi-layer all-optical spiking neural networks can be implemented.